\documentclass{article}

\usepackage[dvips]{color}

\usepackage{natbib}
\usepackage{setspace}
\usepackage{amsmath}
\usepackage{amsmath,epsfig,epsf,psfrag}
\usepackage{verbatim}
\usepackage{rotating}
\usepackage{multirow}
\usepackage{longtable}
\usepackage{booktabs}
\usepackage{subfigure}
\usepackage{lineno}
\usepackage{bm}

\begin{document}


\title{Hazard function models to estimate mortality rates affecting fish populations with application to the sea mullet ({\it Mugil cephalus}) fishery on the Queensland coast (Australia)}

\author{Marco Kienzle\footnote{Queensland Dept of Agriculture, Fisheries, Ecosciences Precinct, Joe Baker St, Dutton Park, Brisbane, QLD 4102, Australia; \newline University of Queensland, School of Agriculture and Food Sciences, St. Lucia, QLD 4072, Australia}}

\maketitle

\section*{Summary}
Fisheries management agencies around the world collect age data for the purpose of assessing the status of natural resources in their jurisdiction. Estimates of mortality rates represent a key information to assess the sustainability of fish stocks exploitation. Contrary to medical research or manufacturing where survival analysis is routinely applied to estimate failure rates, survival analysis has seldom been applied in fisheries stock assessment despite similar purposes between these fields of applied statistics. In this paper, we developed hazard functions to model the dynamic of an exploited fish population. These functions were used to estimate all parameters necessary for stock assessment (including natural and fishing mortality rates as well as gear selectivity) by maximum likelihood using age data from a sample of catch. This novel application of survival analysis to fisheries stock assessment was tested by Monte Carlo simulations to assert that it provided un-biased estimations of relevant quantities. The method was applied to data from the Queensland (Australia) sea mullet ({\it Mugil cephalus}) commercial fishery collected between 2007 and 2014. It provided, for the first time, an estimate of natural mortality affecting this stock: 0.22 $\pm$ 0.08 year$^{-1}$.

\section*{Keywords}
Monte Carlo; natural mortality estimate; survival analysis; fish stock assessment

\section{Introduction}

One purpose of stock assessment is to estimate mortality rates affecting fish stocks. This estimation problem is easier to solve for species that can be aged as opposed to those for which age can't be determined, for example crustaceans. The reason is that mortality and longevity are inversely related, hence age is a measure of mortality. The central mortality model in fisheries research relating catch to the number of fish belonging to a cohort through time was proposed by Baranov \citep{quin99b}. Given recruitment and mortality rates, the proportions of individuals at age in the catch can be calculated and used in a multinomial likelihood  \citep{Four82a}. This method has become by far the most common likelihood to integrate age data into modern stock assessment models \citep{Francis201470, Maunder201361}.\\

The deterministic exponential model in Baranov's catch equation has a statistical counterpart in the form of the exponential probability distribution function which first and second moments quantify the relationship between longevity and mortality rate \citep{cow98b}: the mean age of a cohort which abundance declines at a constant rate is the inverse of that rate. Adopting such a statistical view of the exponential decay of individuals belonging to a cohort allowed the development of a set of maximum likelihood functions to estimate parameters of importance when assessing stocks. The field of survival analysis in statistics has created both a conceptual framework and refined methods to estimate mortality rates \citep{kleinbaum2005survival,cox84b} which are widely applied in medical research and engineering. \\ 

Despite the common goal of estimating survival rates in medical and fisheries research, survival analysis has seldom been applied to stock assessment. \cite{scimar42} proposed to use the Weibull distribution as the survivor function to model data from scientific fishing surveys. In this manuscript, we developed an alternative application of survival analysis to model data from samples of commercial catches using hazard functions derived from time series of fishing effort and a schedule for gear selectivity. Likelihood functions of age data were derived to estimate a constant natural mortality rate, catchability and age-specific gear selectivity. This manuscript starts with a simplistic example using constant natural and fishing mortality rates to introduce fundamental concepts from survival analysis applied to fishery research, before moving to more sophisticated cases leading to an application to real data from the sea mullet fishery in Queensland (Australia). The proposed methods were tested by Monte Carlo, using simulated data sets to characterize some of their properties and assert their capacity to estimate population dynamic parameters of interest to stock assessment. Finally, an application to the mullet fishery case study provided specific estimates of natural mortality, catchability and selectivity. \\

\section{Materials and methods} 

Each fish can be assigned an age by examining its otolith, which is found just below its brain. Fish otoliths deposit calcium carbonate through time, thus increasing in size each year of their life. Microscopic observation of otolith sections often reveal alternate opaque and translucent zones, which can be used to assign individual fish to a particular age group. \\ 

Sampling programs in fisheries research centers around the world aim to collect a representative sample of fish each year to determine the distribution of age of any species of interest. In most cases, the data are binned into age-groups of width 1 year. For this reason, we split the lifespan of cohorts from their birth ($t \in [0;\infty]$) into $n$ yearly intervals from $a_{1}=0$ to the maximum age of $a_{n+1}$ years. While the theory presented here used that particular subdivision of time ($t$), unequal ones also applies. 
 
\subsection{The likelihood for constant natural and fishing mortality rates} 

The exponential decrease in abundance of individuals belonging to a single cohort due to constant natural ($M$) and fishing ($F$) mortality rates was described from a survival analysis point of view \citep{scimar42,cox84b} using a constant hazard function of time ($t$) and parameters $\bm{\theta}$

\begin{equation}
h(t; \bm{\theta}) = M + F
\end{equation}

The probability density function (pdf) describing survival from natural and fishing mortality is

\begin{equation}
f(t; \bm{\theta}) = (M + F) \ e^{-(M+F)t} = \underbrace{M \times e^{-(M+F)t}}_{=f_{1}(t; \bm{\theta})} + \underbrace{F \times e^{-(M+F)t}}_{=f_{2}(t; \bm{\theta})}
\end{equation}

Since age data belonging to individuals dying from natural causes (note that contrary to human, fish's largest cause of natural mortality is to be eaten by another fish) are generally not available to fisheries scientists, we used only the component of the pdf that relates to fishing mortality ($f_{2}(t; \bm{\theta})$). This component of $f(t; \bm{\theta})$ integrates over the entire range of $t$ to 

\begin{align}
\begin{split}
 \int_{t=0}^{t=\infty} f_{2}(t; \bm{\theta}) \ dt &= \int_{t=0}^{t=\infty} F \times e^{-(M+F)t} \ dt\\
                                         &= \int_{t=0}^{t=\infty} f(t; \bm{\theta}) \ dt - \int_{t=0}^{t=\infty} M \times e^{-(M+F)t} \ dt\\
                                         &= 1 - \int_{t=0}^{t=\infty} M \times e^{-(M+F)t} \ dt \\
                                         &= 1 - \frac{M}{M+F}
\end{split}
\end{align}

Hence, the pdf of catch at age data was obtained by normalizing $f_{2}(t; \bm{\theta})$

\begin{align}
\begin{split}
g(t; \bm{\theta}) &= \frac{1}{1 - \frac{M}{M+F}} \ f_{2}(t; \bm{\theta}) \\
             &= \frac{M+F}{F} \ F \times e^{-(M+F)t} \\
             &= f(t; \bm{\theta})
\end{split}
\end{align} 

The likelihood \citep{edwards1992likelihood} of a sample of fish caught in the fishery ($S_{i}$) was written as 

\begin{align}
\begin{split}
\mathcal{L}  &= \prod_{i=1}^{n} \bigl ( \int_{t=a_{i}}^{t=a_{i+1}} f(t; \bm{\theta}) \ dt \bigr ) ^ {S_{i}} \\
            &= \prod_{i=1}^{n} P_{i} ^ {S_{i}}
\end{split}
\end{align}



\noindent where $P_{i}$ is the probability of dying in the interval $[a_{i}; a_{i+1}]$.\\

The logarithm of the likelihood was
\begin{align}
\begin{split}
{\rm log}(\mathcal{L}) &= \sum_{i=1}^{n} S_{i} \ {\rm log} \bigl ( \int_{t=a_{i}}^{t=a_{i+1}} f(t; \bm{\theta}) \ dt \bigr ) \\
                       &= \sum_{i=1}^{n} S_{i} \ {\rm log} \bigl ( \int_{t=a_{i}}^{t=a_{i+1}} (M + F) \ e^{-(M+F)t} \ dt \bigr ) \\
                       &= \sum_{i=1}^{n} S_{i} \ {\rm log} \bigl ( e^{-(M+F) \times a_{i}} - e^{-(M+F) \times a_{i+1}} \bigr )
\label{eq:16}
\end{split}
\end{align}

The log-likelihood can accommodate a last age-group made of all observations above a certain age in the sample (referred to as a +group) as follow \citep{pawitan2013all}

\begin{equation}
{\rm log}(\mathcal{L}) = \sum_{i=1}^{n-1} S_{i} \ {\rm log} \bigl ( e^{-(M+F) \times a_{i}} - e^{-(M+F) \times a_{i+1}} \bigr ) + S_{n} \ {\rm log} \bigl ( e^{-(M+F) \times a_{n}} \bigr )
\end{equation}



This development illustrated an application of survival analysis to estimate mortality rates affecting a cohort of fish by maximum-likelihood using a sample of catch at age. This method was implemented in R \citep{R} in the package Survival Analysis for Fisheries Research (SAFR).

Natural and fishing mortality cannot be disentangled with catch data only but the next section will show that the provision of effort data allowed to estimate both catchability ($q$) and natural mortality.

\subsection{Estimating catchability and natural mortality}

In this section, we assumed that a time series of effort ($E_{i}$) associated with a sample of catch at age ($S_{i}$) was available to the researcher. And the assumption that fishing mortality varied according to fishing effort through constant catchability ($q$) held: $F(t) = q \ E(t)$. In this situation, the hazard function was written as

\begin{equation}
h(t,\bm{\theta}) = M + q \ E(t)
\end{equation} 

And the pdf

\begin{align}
\begin{split}
f(t, \bm{\theta}) &= ( M + q \ E(t)) \ e^{-Mt-q\int_{0}^{t}E(t) \ dt} \\
             &= \underbrace{M \times e^{-Mt-q\int_{0}^{t}E(t) \ dt}}_{=f_{1}(t; \bm{\theta})} + \underbrace{q \ E(t) \times e^{-Mt-q\int_{0}^{t}E(t) \ dt}}_{=f_{2}(t; \bm{\theta})}
\end{split}
\end{align}

As in the previous section, we had

\begin{equation}
\int_{t=0}^{t=\infty} f_{2}(t; \bm{\theta}) \ dt = 1 - \int_{t=0}^{t=\infty} M \times e^{-Mt-q\int_{0}^{t}E(t) \ dt} \ dt \\
\end{equation}

But we did not know an analytic solution to the integral since the function $E(t)$ was not specified. Nevertheless, given effort in every interval ($\int_{t=a_{i}}^{t=a_{i+1}}  E(t) \ dt = E_{i} = \int_{t=0}^{t=a_{i+1}} E(t) \ dt - \int_{t=0}^{t=a_{i}} E(t) \ dt, \forall i \in [1; n]$), we could calculate the value of $\int_{t=0}^{t=\infty} f_{2}(t; \bm{\theta}) \ dt$. 

\begin{align}
\begin{split}
\int_{t=0}^{t=\infty} f_{2}(t; \bm{\theta}) &= 1 - \sum_{i=1}^{n} \bigl [ -\frac{M}{M+q \ E_{i}} e^{-Mt-q\int_{0}^{t}E(t) \ dt} \bigr ]_{t=a_{i}}^{t=a_{i+1}} \\
                                   &= 1 - \sum_{i=1}^{n} \frac{M}{M+q \ E_{i}} \bigl ( e^{-M \ a_{i}-q \int_{0}^{a_{i}}E(t) \ dt} - e^{-M \ a_{i+1}-q \int_{0}^{a_{i+1}} E(t) \ dt} \bigr ) \\
                                   &= \sum_{i=1}^{n} \frac{q \ E_{i}}{M+q \ E_{i}} \bigl ( e^{-M \ a_{i}-q\int_{0}^{a_{i}}E(t) \ dt} - e^{-M \ a_{i+1}-q\int_{0}^{a_{i+1}}E(t) \ dt} \bigr ) \\
\end{split}
\end{align}

In practice, $\int_{t=0}^{t=\infty} f_{2}(t; \bm{\theta})$ is bound between 0 and 1. It takes a specific value depending on the values of $M, q $ and $E_{i}$. Naming this constant value $K$, we could write the pdf of catch at age given effort data are available as

\begin{equation}
g(t; \bm{\theta}) = \frac{1}{K} \ f_{2}(t; \bm{\theta})
\end{equation}

And the log-likelihood:
\begin{equation}
{\rm log}(\mathcal{L}) = \sum_{i=1}^{n} S_{i} \ {\rm log} \bigl ( \int_{t=a_{i}}^{t=a_{i+1}} g(t; \bm{\theta}) \ dt \bigr )
\end{equation}

Accounting for age-specific gear selectivity ($s(t)$) effects on fishing mortality ($F(t) = q \ s(t) \ E(t)$) was included in a similar way into the likelihood using constant value for selectivity at age. In practice, it is difficult to estimate $n$ additional selectivity parameters using only the data from a single cohort but processing several cohorts at the same time assuming separability of fishing mortality rendered estimation of catchability, natural mortality and selectivity possible.

\subsection{Estimates from catch at age matrix using fishing mortality separability}

This section describes an application of survival analysis to matrices of catch at age, developed for the purpose of estimating catchability ($q$), selectivity at age ($s(t)$) and constant natural mortality ($M$). The matrix ($\bm{S}_{i,j}$) containing a sample of fishes aged to belong to a particular age-group $j$ in year $i$ contains $n+p-1$ cohorts. These cohorts were indexed by convention using $k$ ($k \in [1, n+p-1]$) and an increasing number $r_{k}$ ($ 1 \leq r_{k} \leq {\rm min}(n,p)$) identifying incrementally each age-group (see Appendix p.~\pageref{Appendix:DefinitionsOfMathematicalSymbols} for more information). Each matrix $\bm{S}_{i,j}$ has two cohorts with only 1 age-group representing them.\\

The derivation for a single cohort were the same as those presented in the previous section, here reproduced with indexations relative to a single cohort and accounting for selectivity

\begin{equation}
g_{k}(t; \bm{\theta}) = \frac{q \ s(t) \ E(t) \times e^{-Mt-q\int_{0}^{t} s(t) \ E(t) \ dt}}{\sum_{l=1}^{r_{k}} \frac{q \ s_{k,l} \ E_{k,l}}{M+q \ s_{k,l} \ E_{k,l}} \bigl ( e^{-M \ a_{k,l}-q\int_{0}^{a_{k,l}}s(t) \ E(t) \ dt} - e^{-M \ a_{k,l}-q\int_{0}^{a_{k,l+1}}s(t) \ E(t) \ dt} \bigr )} 
\end{equation}
\newline
The likelihood function of a catch at age matrix was built using each pdf specific to each cohort ($g_{k}(t; \bm{\theta})$):

\begin{equation}
\mathcal{L} = \prod_{k=1}^{n+p-1} \prod_{l=1}^{r_{k}}  \bigl ( \int_{t=a_{k,l}}^{t=a_{k,l+1}} g_{k}(t; \bm{\theta}) \ dt \bigr ) ^ {\bm{S}_{k,l}}
\end{equation}

The expression above is equivalent to 
\begin{equation}
\mathcal{L} = \prod_{i,j} \bm{P}_{i,j} ^ {\bm{S}_{i,j}}
\end{equation}

\noindent where the $\bm{P}_{i,j}$ are the probabilities of observing a fish of a given age $j$ in year $i$ given by the hazard model. In this likelihood, the $\bm{P}_{i,j}$ sum to 1 along the cohort instead of summing to 1 for each year as described for the multinomial likelihood in \cite{Four82a}. \\

\subsection{Monte Carlo simulations}

The method presented in the previous section to estimate mortality and selectivity from a matrix containing a sample of number at age were tested with simulated data sets to characterize their performance. Variable number of cohorts ($n+p-1 = 25$, 35 or 45); maximum age ($p=8$, 12 or 16 years) and sample size of age measurement in each year varying from 125 to 2000 increasing successively by a factor 2 were used. The simulated data sets were created by generating an age-structure population using random recruitment for each cohort, random constant natural mortality, random catchability and random fishing effort in each year (Tab.~\ref{tab:SimulationParameters.tex}). A catch at age matrix was calculated using a logistic gear selectivity with 2 parameters: 

\begin{equation}
\bm{s}_{a_{i}} = \frac{1}{1  + {\rm exp}( \alpha - \beta \times a_{i})}
\end{equation}


\begin{table}[ht]
\centering
\begin{tabular}{lll}
  \hline
Variable type & Distribution & Parameters \\
  \hline
recruitment               & uniform & min=$10^{6}$, max=$10^{7}$   \\
natural mortality         & uniform & min=0.1, max=0.8 \\
catchability              & uniform & min=$10^{-4}$, max=$10^{-3}$ \\
fishing effort            & uniform & min=$10^{3}$, max=$5 \times 10^{3}$   \\
gear selectivity $\alpha$ & uniform & min=8, max=12      \\
gear selectivity $\beta$  & uniform & min=1, max=3       \\
   \hline
\end{tabular}
\caption{Distribution and range of value taken by different type of random variable in simulations.}
\label{tab:SimulationParameters.tex}
\end{table}

Several sampling strategies were implemented to assess how it affected mortality estimates. To test estimators derived from survival analysis, one would like to draw randomly from the probability distribution. This is obviously impossible in the real world because field biologists never have in front of them a entire cohort to chose from. Nevertheless, we implemented a sampling strategy (sampling strategy 1) that randomly selected from the entire simulated catch at age matrix ($\bm{C}_{i,j}$) as a benchmark. In the real world, samples can be drawn by accessing only a single year-class of every cohort every year, so the second strategy implemented was to simulate a random selection of a fixed number of sample ($N$) each year (sampling strategy 2). Finally, the third strategy investigated (sampling strategy 2 with weighting) was to apply a weighting by the estimated total catch at age ($\hat{\bm{C}}_{i,j}$) to the sample of number at age in the sample ($\bm{S}_{i,j}$):

\begin{equation}
\hat{\bm{C}}_{i,j} = \bm{p}_{i,j} \odot \bm{C}_{i} \otimes \bm{v}(j)
\end{equation}

\noindent where $\bm{p}_{i,j}$ is the proportion at age (see Appendix p.~\pageref{Appendix:DefinitionsOfMathematicalSymbols}), $\bm{C}_{i}$ is a column vector containing the total number of fish caught in each year $i$ and $\bm{v}(j)$ is a row vector of 1's. A weighted sample ($\bm{S}^{*}_{i,j}$) was obtained using the fraction of total catch sampled
\begin{equation}
\bm{S}^{*}_{i,j} = \hat{\bm{C}}_{i,j} \times \frac{\sum_{i,j} \bm{S}_{i,j}}{\sum_{i} \bm{C}_{i}}
\end{equation}

Note that $\sum_{i,j} \bm{S}_{i,j} = \sum_{i,j} \bm{S}^{*}_{i,j}$.\\


Comparisons with the multinomial likelihood proposed by \cite{Four82a} were made using differences in negative log-likelihood between that method and the hazard function approach described in the present article. Simulated catch were used to calculate the proportion of individual at age, constraining them to sum to 1 in each year. This method to calculate proportions for the multinomial likelihood was regarded as the best case scenario because we expect any estimation algorithm based on the multinomial likelihood to, at best, match exactly the simulated catch at age. The logarithm of these proportions were then multiplied by the simulated age sample (weighted or not depending on the case) to calculate the log-likelihood as described in \cite{Four82a}. This quantity was compared to that calculated using the survival analysis approach to determine which model best fitted the simulated data. This comparison ignored the number of parameters used in each model. The multinomial likelihood requires $n+p-1$ more parameters to be estimated than the survival analysis because the former requires an estimate of recruitment for each cohort in order to calculate the proportion at age in the catch.

\subsection{A case study: Queensland's sea mullet fishery}

The straddling sea mullet ({\it Mugil cephalus}) is caught along the east coast of Australia, with most landings occurring between 19$^{o}$S (approx. Townsville) and 37$^{o}$S (roughly the border between New South Wales and Victoria). The most noticeable feature of the biology of this species is a massive northward spawning migration of the stocks along the coast during autumn \citep{Kesteven53a}. Tagging experiments revealed that 90\% of tagged animals travelled less than 85 km during the migration season \citep{Kesteven53a}. Analyses of parasites concluded that the bulk of sea mullet caught in Queensland fishery is based on local fish populations and not migrating from New South Wales \citep{Lester2009129}. Following recommendations from \cite{Bell2005r}, an existing (1999--2004) scientific survey design was modified from 2007 onward to include both estuaries and ocean habitats in order to provide representative demographic statistics of the fish caught in Queensland. Samples were collected in both habitats (Tab.~\ref{tab:Mullet-NbAtAge}). Age varied between 0 and 17 years. A 14+ age-group was created to combine the small number of observations in the older age-groups. These data were weighted by catch in each year and habitat to obtain a dataset representative of the entire catch in this fishery. \\

\begin{sidewaystable}[ht]
\tiny
\centering
\begin{tabular}{|l|c|c|c|c|c|c|c|c|c|c|c|c|c|c|c|c|c||c||c|}
\hline
& \multicolumn{19}{|c|}{Estuaries} \\
\hline
 & 0--1 & 1--2 & 2--3 & 3--4 & 4--5 & 5--6 & 6--7 & 7--8 & 8--9 & 9--10 & 10--11 & 11--12 & 12--13 & 13--14 & 14--15 & 15--16 & 16--17 & Catch & Effort \\ 
  \hline
2007 & 8 & 144 & 298 & 233 & 23 & 44 & 17 & 6 & 2 &  &  &  &  &  &  &  &  & 792 & 6834 \\ 
  2008 &  & 25 & 242 & 265 & 144 & 42 & 22 & 4 & 2 &  &  &  &  & 1 &  &  &  & 1089 & 7228 \\ 
  2009 &  & 85 & 131 & 332 & 88 & 39 & 9 & 3 & 1 & 1 & 1 &  &  & 1 &  &  &  & 950 & 6045 \\ 
  2010 & 2 & 180 & 306 & 133 & 87 & 29 & 19 & 3 & 4 &  &  &  &  &  &  &  &  & 827 & 5640 \\ 
  2011 & 4 & 176 & 409 & 236 & 38 & 15 & 11 & 5 &  & 1 &  &  &  &  & 1 &  &  & 737 & 5852 \\ 
  2012 & 1 & 83 & 437 & 253 & 108 & 23 & 8 & 3 & 4 &  &  &  &  &  &  &  &  & 938 & 6527 \\ 
  2013 &  & 76 & 290 & 515 & 250 & 73 & 10 & 6 & 1 &  &  &  &  &  &  &  &  & 1152 & 6083 \\ 
  2014 & 2 & 47 & 211 & 227 & 186 & 78 & 18 & 5 &  &  &  &  &  &  &  &  & 1 & 645 & 4777 \\ 
\hline
& \multicolumn{19}{|c|}{Ocean} \\
\hline
  2007 & 3 & 36 & 219 & 328 & 95 & 61 & 28 & 18 & 9 & 3 &  &  &  &  &  &  &  & 559 & 566 \\ 
  2008 &  & 17 & 226 & 353 & 265 & 58 & 35 & 17 & 8 & 8 & 2 & 2 &  & 1 &  &  &  & 706 & 647 \\ 
  2009 & 1 & 25 & 149 & 347 & 163 & 112 & 20 & 14 & 5 &  & 2 & 2 & 1 &  &  &  &  & 865 & 484 \\ 
  2010 &  & 59 & 235 & 117 & 113 & 68 & 31 & 8 & 5 & 2 &  &  &  &  &  &  &  & 930 & 469 \\ 
  2011 & 2 & 68 & 189 & 264 & 77 & 56 & 24 & 5 & 2 & 3 &  & 1 & 1 &  &  & 1 &  & 805 & 560 \\ 
  2012 &  & 16 & 196 & 310 & 190 & 34 & 24 & 12 & 7 &  &  &  &  &  &  &  &  & 711 & 467 \\ 
  2013 &  & 13 & 115 & 440 & 282 & 110 & 15 & 18 & 4 &  &  &  &  & 1 &  &  &  & 857 & 578 \\ 
  2014 &  & 40 & 139 & 232 & 355 & 131 & 40 & 6 & 4 &  &  &  &  &  &  &  &  & 813 & 441 \\ 
   \hline
\end{tabular}
\caption{Distribution of yearly samples (in rows) of sea mullet into age-groups of width 1 year (in columns) in the estuaries and ocean habitats; catch in tonnes and effort in number of days.} 
\label{tab:Mullet-NbAtAge}
\end{sidewaystable}

Sea mullet are thought to spawn in oceanic waters adjacent to ocean beaches from May to August each year. By convention, the birth date was assumed to be on July 1$^{st}$ each year. Opaque zones are thought to be deposited on the otolith margin during spring through early summer (September to December). Biologists have come to the conclusion that the first identifiable opaque zone is formed 14 to 18 months after birth, and all subsequent opaque zones are then formed at a yearly schedule \citep{Smith2003}. Each fish in the sample was assigned an age-group based on opaque zone counts and the amount of translucent material at the margin of otolith. Age-group 0--1 comprised fish up to 18 months old ($a_{1}=18$ months) while all subsequent age-groups spanned 12 months ($a_{2} = 30$ months, $a_{3}= 42$ months, etc ...).\\

Three hazard function models were fitted to the data: a first model assumed a constant natural mortality across age-groups and throughout the period covered by the data, a common catchability and gear selectivity in estuaries and ocean (model 1, Tab.~\ref{tab:MulletModelComparison}); the second model assumed that catchability differed between estuaries and ocean; and the third model assumed that both catchability and gear selectivity differed between the two habitats. The models were compared using Akaike Information Criteria (AIC) to determine which was most supported by the data \citep{Burnb03}.\\

\begin{table}[ht]
\centering
\begin{tabular}{rrrr}
  \hline
Model & $p$ & $-{\rm log}(\mathcal{L})$ & AIC \\ 
  \hline
   2 & 17 & 18795.3 & 37624.6 \\ 
   3 & 31 & 18787.7 & 37637.4 \\ 
   1 & 16 & 18817.7 & 37667.4 \\ 
   \hline
\end{tabular}
\caption{Comparison between the log-likelihood value obtained for several hazard function models of the mullet data using different numbers of parameters ($p$). The models were ordered by increasing values of Akaike Information Criteria (AIC =$-2 {\rm log}(\mathcal{L}) + 2p$) from top to bottom.} 
\label{tab:MulletModelComparison}
\end{table}


\section{Results}

    \subsection{Method tests using simulated data}

\begin{figure}
\psfrag{dsd1}[t][t]{{\protect\large $\displaystyle\frac{ds}{d\phi_1}$}}
\psfrag{phi1}[t][t]{{\protect\Large $\phi_1$}}
\centering
\epsfig{file=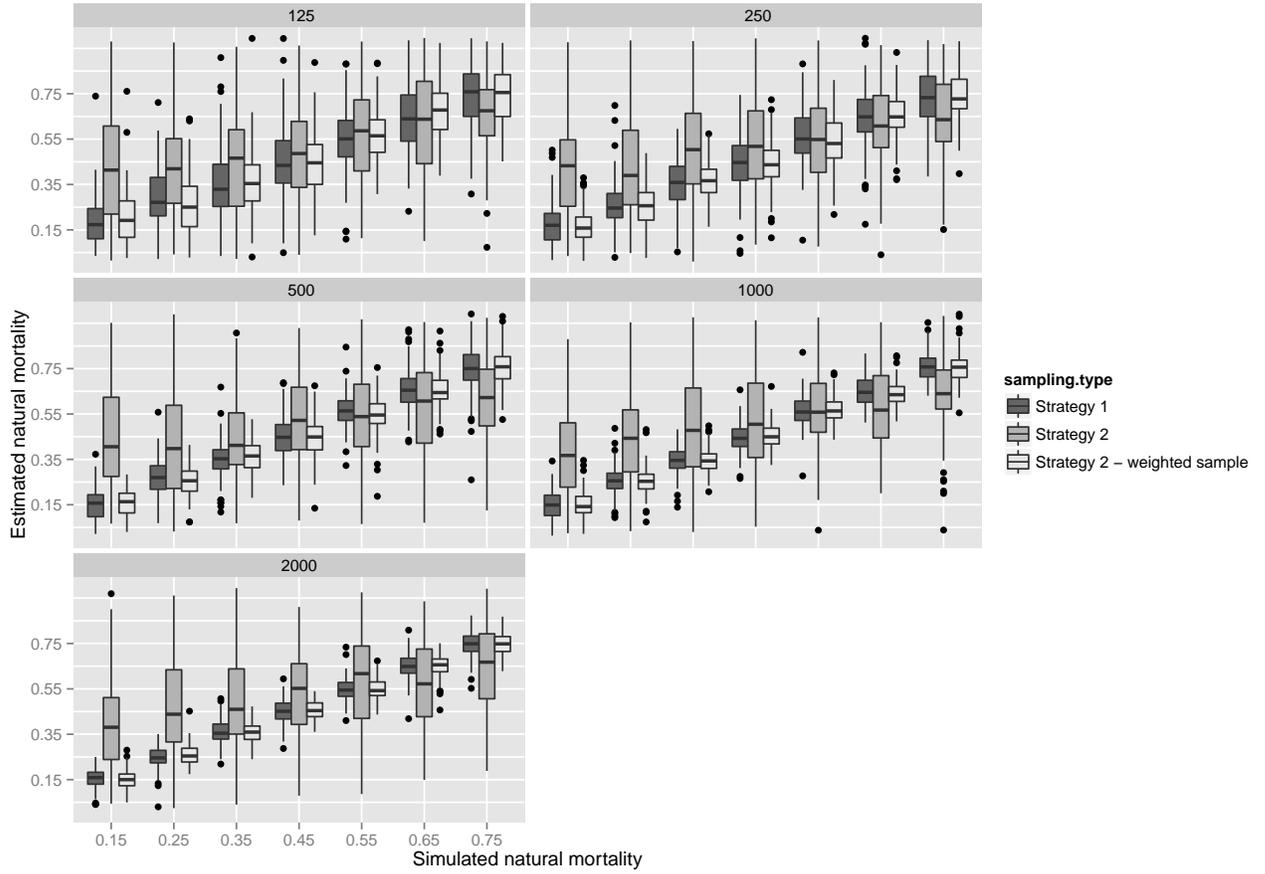,width=12cm,angle=-90}
\vspace{2cm}
\caption{Comparison between simulated natural mortality (x-axis) and estimated (y-axis) using a random sample of the matrix of catch (strategy 1); a random sample from each year separately (strategy 2) and the same sample weighted by yearly total catch (strategy 2 - weighted samples). Each panel correspond to an increasing number of samples per year varying from 125 to 2000.}
\label{fig:Estimating-NaturalMortality}
\end{figure}

\begin{figure}
\psfrag{dsd1}[t][t]{{\protect\large $\displaystyle\frac{ds}{d\phi_1}$}}
\psfrag{phi1}[t][t]{{\protect\Large $\phi_1$}}
\centering
\epsfig{file=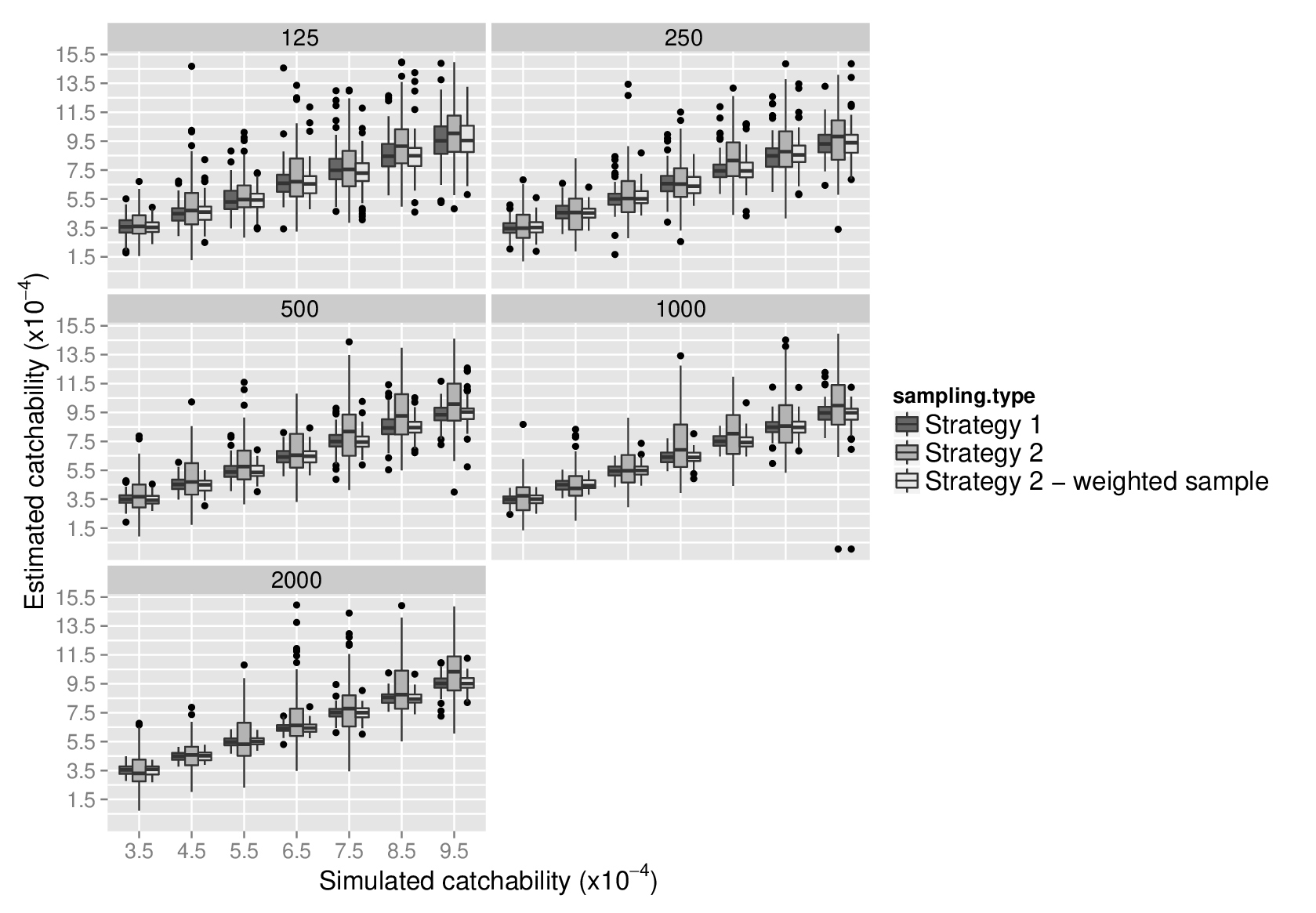,width=12cm,angle=-90}
\vspace{2cm}
\caption{Comparison between simulated catchability (x-axis) and estimated (y-axis) using a random sample of the matrix of catch (strategy 1); a random sample from each year separately (strategy 2) and the same sample weighted by yearly total catch (strategy 2 - weighted samples). Each panel correspond to an increasing number of samples per year varying from 125 to 2000.}
\label{fig:Estimating-Catchability}
\end{figure}

Weighting the numbers of sampled fish each year by total catch (sampling strategy 2 - weighted sample) performed as well as the benchmark sampling strategy 1 (Fig.~\ref{fig:Estimating-NaturalMortality} and Fig.~\ref{fig:Estimating-Catchability}). By contrast, estimations using a fixed number of fish each year were biased suggesting that weighting by catch is necessary in practical applications of the survival analysis approach. \\

Weighting of age-data samples considerably reduced the variability of natural mortality estimates (Fig~\ref{fig:Estimating-NaturalMortality}). Increasing the number of samples reduced uncertainty associated with natural mortality estimates too. \\ 

Estimates of catchability were much more consistent across the range of values tested (3--10 $ \times 10^{-4}$) for all methods (Fig.~\ref{fig:Estimating-Catchability}). The bias of the unweighted approach (strategy 2) was often similar to that of the weighted one (strategy 2 - weighted sample). But the uncertainty associated with the former approach was much larger than the latter. For both strategy 1 and strategy 2 with weighting, the benefit of increasing sampling size were very noticeable up to a 1000 fish aged but less so beyond that.\\

\begin{figure}
\psfrag{dsd1}[t][t]{{\protect\large $\displaystyle\frac{ds}{d\phi_1}$}}
\psfrag{phi1}[t][t]{{\protect\Large $\phi_1$}}
\centering
\epsfig{file=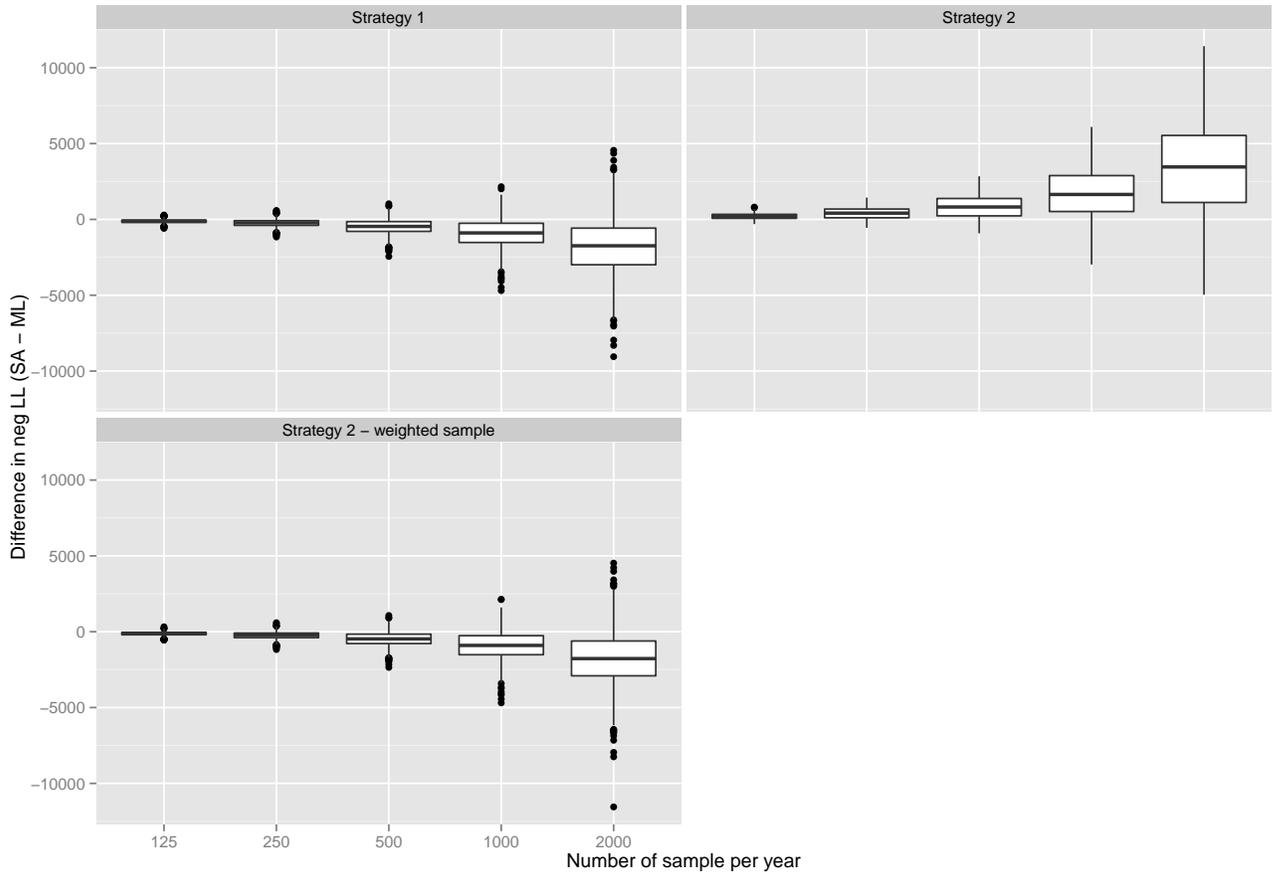,width=12cm,angle=-90}
\vspace{2cm}
\caption{Difference between the negative log-likelihood ($-{\rm log}(\mathcal{L})$) from survival analysis (SA) and multinomial (ML) as a function of the number of sample per year. Each panel represents a particular sampling strategy.}
\label{fig:ComparisonOfNegLL}
\end{figure}

The comparison between the likelihood function from survival analysis and the multinomial likelihood (Fig.~\ref{fig:ComparisonOfNegLL}) showed that, apart sampling strategy 2 which provided biased estimates, the approach using survival analysis provided in the majority of cases smaller negative log-likelihood values than the multinomial likelihood. The substantial advantage given the multinomial likelihood in this comparison played an important role at low sampling intensity where the assumption that proportion at age was known perfectly artificially improved its performance in most difficult situations. This artificial advantage faded away as the simulated sample sizes were increased resulting in the survival analysis approach outperforming the multinomial likelihood. \\

    \subsection{Mortality estimates for sea mullet}

Sea mullet data showed larger catch per unit of effort in the ocean than in estuaries (Tab.~\ref{tab:Mullet-NbAtAge}). Of all three models compared with AIC, the model that assumed catchability varied between habitats and selectivity was the same in both habitats (model 2) was best supported by the data (Tab.~\ref{tab:MulletModelComparison}). This model estimated catchability in the ocean to be 16 times larger than in estuaries (Tab.~\ref{tab:BestParameterEstimates}). Natural mortality for sea mullet was estimated to be equal to 0.219 $\pm$ 0.082 year$^{-1}$. Estimates of gear selectivity suggested it increased up to the fifth age-group, beyond which fishes were fully selected by the fishing gear.\\

\begin{table}[ht]
\centering
\begin{tabular}{ll}
  \hline
Parameters & Estimates \\ 
  \hline
q$_{\rm{estuaries}}$ & (4.527 $\pm$ 1.026) $\times 10^{-5}$\\ 
  q$_{\rm{ocean}}$ & (7.243 $\pm$ 1.142) $\times 10^{-4}$\\ 
  M & 0.219 $\pm$ 0.082 \\ 
  $s_{1}$ & 0.002 $\pm$ $2 \times 10^{-6}$ \\ 
  $s_{2}$ & 0.08 $\pm$ 0.011 \\ 
  $s_{3}$ & 0.374 $\pm$ 0.016 \\ 
  $s_{4}$ & 0.771 $\pm$ 0.045 \\ 
  $s_{5}$ & 0.993 $\pm$ 0.07 \\ 
  $s_{6}$ & 1.000 $\pm$ 0.056 \\ 
  $s_{7}$ & 1.000 $\pm$ 0.021 \\ 
  $s_{8}$ & 1.000 $\pm$ 0.111 \\ 
  $s_{9}$ & 1.000 $\pm$ 0.195 \\ 
  $s_{10}$ & 0.905 $\pm$ 0.326 \\ 
  $s_{11}$ & 0.904 $\pm$ 0.503 \\ 
  $s_{12}$ & 0.905 $\pm$ 0.389 \\ 
  $s_{13}$ & 0.905 $\pm$ 0.528 \\ 
  $s_{14}$ & 1.000 $\pm$ 0.661 \\ 
   \hline
\end{tabular}
\caption{Maximum likelihood parameters estimates from model 2.}
\label{tab:BestParameterEstimates}
\end{table}


\section{Discussion}

This application of survival analysis to fisheries research provided an effective approach to develop maximum likelihood estimators of natural and fishing mortality rates, and gear selectivity, from age data. Monte Carlo simulations showed that it provided unbiased estimates of natural mortality and catchability over a wide range of simulated values. \\

The comparison between the negative log-likelihood from the survival analysis approach with the multinomial likelihood \citep{Four82a} suggested that the former offered a better model to represent the data. This comparison was made using the best possible outcome for the multinomial likelihood because it used the simulated proportions of individuals at age in place of the probabilities to compute the likelihood. Arguably, a substantial advantage was given to the multinomial likelihood over the survival analysis in this comparison because no one would reasonably expect any estimation method to systematically provide exactly the proportion at age in the catch using a sample of the data. Therefore, the present comparison really focused on which probabilities to use in the likelihood function, whether they should sum to 1 in each year along age-groups or along cohorts. Despite the strong advantage given to the multinomial likelihood, the results suggested that simulated data according to Baranov's catch equation were fundamentally better fitted by a statistical method that modelled the exponential decay of individuals along cohorts rather than by one that assumed the data followed a multinomial probability distribution specific to each year.\\

Weighting of the sample provided unbiased estimates of natural mortality and catchability. Mortality estimates, in particular fishing mortality, depended on the magnitude of catch. The unrealistic sampling strategy which assumed that all catch data would be in front of the experimenter at once for sampling, accounted automatically for variation of catch and effort in each year because the abundance of each age-group in the catch determined the probability to choose at random an individual belonging to any age-group. In practical application of survival analysis to fishery research, weighting is necessary because one cannot know {\it a priori} the magnitude of catch in coming years. \\


The Monte Carlo simulations used a logistic gear-selectivity to generate and fit the data although we would have preferred to generate data from a wide range of possible gear-selectivity functions or even using non-parametric procedures. Simulations showed that gear selectivity were the most difficult parameters to estimate. The sea mullet case study was in fact not fitted with a logistic curve but selectivity were estimated through a tedious process to search each proportion retained at age that best fitted the data as measured by the likelihood. This process could not be automatized into the simulation testing framework to provide automatic identification of gear-selectivity. This aspect of the analysis was left out of the present manuscript for future work. Criticisms that this somewhat simplified the problem would be justified. But the current article was designed as an introduction to the application of survival analysis to fisheries catch at age data, not one that solves all problems at once. As such, the likelihood approach presented in this manuscript provides a method to identify the gear selectivity that best fit the data, just not an automatic one. \\ 



The model best supported by the mullet data set estimated natural mortality equal to $0.22 \pm 0.08$. This is the first estimate of natural mortality for mullet in Australia. Previous to this estimation, it was customary to use the natural mortality estimated by linear regression from \cite{Hwang82a} for the mullet fishery in Taiwan (M=0.33 year$^{-1}$) which fall within 2 S.D. of the estimate for the Queensland fishery. The model that fitted best the mullet data estimated catchability in the ocean to be 16 times larger than in estuaries. This is consistent with fishermen reporting very large catches from their ocean beach operations (up to 40 tonnes per haul) compared to working in estuaries.\\




This likelihood method may well find its place naturally into integrated stock assessment \citep{Maunder201361} as it provided an efficient method to deal with samples of age data. Applications of survival analysis to fishery data could be expanded further. A particular area of interest would be to derive recruitment estimates using the probabilities estimated by survival analysis and total catch from the fishery.

\section*{Acknowledgements}







\clearpage
\newpage
\section{Appendix}

\subsection{Definitions of some mathematical symbols}
\label{Appendix:DefinitionsOfMathematicalSymbols}

This appendix contains definitions of some of the mathematical symbols used in previous sections

\begin{itemize}

\item $\bm{S}_{i,j}$: a matrix of dimensions $n \times p$ ($i \in [1, n]$ and $j \in [1, p]$) containing a number of fishes that were aged and found to belong to specific age-groups $j$ in a particular year $i$. This matrix contains data belonging to $n+p-1$ cohorts, which by convention were labeled using $k$ varying from 1 on the top-right corner of the matrix to $n+p-1$ on the bottom-left (Tab.~\ref{Tab:Cohorts}).

\begin{table}[ht]
\begin{center}

\begin{tabular}{c|cccccc}
 \multicolumn{1}{c}{} & 1 & & \dots & & & $p$ \\
\cmidrule(r){2-7}
 1      & \dots   & \dots & \dots & 3     & 2       & 1      \\
        & \dots   & \dots & \dots & \dots & 3       & 2      \\
\vdots  & \dots   & \dots & \dots & \dots & 4       & 3      \\
        & \dots   & \dots & $k$   & \dots & \dots   & \dots  \\
        & \dots   & \dots & \dots & \dots & \dots   & \dots  \\
$n$     & $n+p-1$ & \dots & \dots & \dots & \dots   & \dots  \\
\end{tabular}

\caption{Convention used to associate each element of the catch at age matrix ($\bm{C}_{i,j}$) with particular cohort referred to as with the number given in this table.}
\label{Tab:Cohorts}

\end{center}
\end{table}

The number of data in $\bm{S}_{i,j}$ belonging to each cohort ($r_{k}$) varies from $1$ to ${\rm min}(n,p)$ and was determined as follow:

\begin{equation}
r_{k} = 
\begin{cases}
i - j + p \ {\rm if} \ k < {\rm min}(n,p)  \\ 
{\rm min}(n,p) \ {\rm if} \ {\rm min}(n,p) \leq k < {\rm max}(n,p) \\
j - i + n \ {\rm if} \ k \geq {\rm max}(n,p) \\
\end{cases}
\end{equation}

\noindent Each element of the $\bm{S}_{i,j}$ matrix is uniquely identified using indices $i$ and $j$ ( $1 \leq i \leq n$ and $ 1 \leq j \leq p$) or indices $k$ and $l$ ( $ 1 \leq k \leq n+p-1 $ and $ 1 \leq l \leq r_{k}$ ), so for example
\begin{equation}
  \sum_{i,j} \bm{S}_{i,j} = \sum_{k,l} \bm{S}_{k,l}
\end{equation}

\item $p_{i,j}$: a matrix of dimensions $n \times p$ ($i \in [1, n]$ and $j \in [1, p]$) containing the proportion at age in the sample ($\bm{S}_{i,j}$). Rows of this matrix sum to 1.

\begin{equation}
\bm{p}_{i,j} = \frac{\bm{S}_{i,j}}{\sum_{j} \bm{S}_{i,j}}
\end{equation}

\item $\bm{F}_{i,j}$ a matrix of fishing mortality with dimension $n \times p$ ($i \in [1, n]$ and $j \in [1, p]$). This matrix was constructed as the outer product of year specific fishing mortality rates ($q \ \bm{E}_{i}$) and selectivity at age ($\bm{s}_{j}$):
\begin{equation}
\bm{F}_{i,j} = q \ \bm{E}_{i} \otimes \bm{s}_{j}
\end{equation}

\end{itemize}

\clearpage
\newpage


\end{document}